\documentclass[10pt, conference, compsocconf]{IEEEtran}
\ifCLASSINFOpdf
\else
\fi
\usepackage{color}
\usepackage{booktabs}

\newcommand{\eg}{{\em e.g.}, }
\newcommand{\ie}{{\em i.e.}, }

\newcommand{\name}{ECCP}

\usepackage[pdftex]{graphicx}
\usepackage[numbers,sort]{natbib}
\usepackage{amsthm}

\begin{document}
%
\title{A Provably-Correct Protocol for Seamless Communication\\ with Mobile, Multi-Homed Hosts}


\author{\IEEEauthorblockN{Matvey Arye, Erik Nordstr{\"o}m, Robert
    Kiefer, Jennifer Rexford, Michael J. Freedman}
\IEEEauthorblockA{Princeton University\\
Princeton, NJ \\
\{arye, enordstr, rkiefer, jrex, mfreed\}@cs.princeton.edu}
}



%


\maketitle

\begin{abstract}
Modern consumer devices, like smartphones and tablets, have multiple
interfaces (e.g., WiFi and 3G) that attach to new access points as users
move.  These mobile, multi-homed computers are a poor match with an
Internet architecture that binds connections to fixed end-points with
topology-dependent addresses.  As a result, hosts typically cannot
spread a connection over multiple interfaces or paths, or change
locations without breaking existing connections.  

In this paper, we introduce {\name}, an end-host connection control
protocol that allows hosts to communicate over multiple interfaces with
dynamically-changing IP addresses.  Each {\name} connection consists of one
or more flows, each associated with an interface or path.  A host can
move an existing flow from one interface to another or change the IP
address using in-band signaling, without any support from the
underlying network.  We use formal models to verify that {\name} works
correctly in the presence of packet loss, out-of-order delivery, and
frequent mobility, and to identify bugs and design limitations in
earlier mobility protocols.
\end{abstract}

\begin{IEEEkeywords}
migration; mobile devices; network architecture; in-band signaling; formal methods;
\end{IEEEkeywords}

%
\IEEEpeerreviewmaketitle

\section{Introduction}
\label{sec:intro} 
The end-to-end argument is a classic design principle of the Internet.  This
simple yet powerful idea---that end-hosts should manage their own communication
without the involvement of intermediaries---was a major factor in the huge
success of the Internet. However, TCP/IP was designed in an era when each
host connected to a single, fixed attachment point.  In contrast, today's
Internet-connected devices have multiple interfaces (\eg WiFi and 3G) and can move
frequently.  To leverage the full capabilities of modern devices, the end-host
protocols should change to support {\em path multiplicity} (where a single
connection is spread over multiple interfaces or paths) and {\em location
  dynamism} (where hosts can change locations without breaking ongoing
connections).

Existing solutions address location dynamism by redirecting traffic
when hosts move.  This requires adding middleboxes (like home agents
in Mobile IP), and leads to inefficient ``triangle routing.''  Other
solutions, like placing multiple wireless access points in the same
virtual LAN (VLAN), support only limited mobility within a single
subnet.  Previous research proposals have proposed flat addressing, to
allow hosts to retain their addresses as they move, at the expense of
new scalability and deployment challenges.  Other work that used
end-to-end connection protocols was either
under-specified~\cite{tcp-r}, thus missing important subtleties, or
exhibited incorrect behavior~\cite{tcp_migrate}.

In this paper, we design a provably correct end-to-end connection control
protocol ({\name}) that supports migration, multiple interfaces, and mobility.
The solution works on top of the IP protocol and location-dependent addresses,
enabling incremental deployment on today's Internet.
Our design rests on an extensive study of existing protocols and
proposals, and a practical experience in end-host stack
design~\cite{serval}. {\name} supports location dynamism by allowing a
device to inform all of its correspondent hosts of the new address.
Path multiplicity is supported by allowing a single connection to
consist of one or more {\em flows}, each associated with an interface
or path. Flows can change to different interfaces or IP addresses over
time, without breaking their connection.

The TCP/IP network stack couples connection control (e.g., starting
and stopping connections and flows, and changing the addresses
associated with flows) with data delivery functionality (e.g.,
congestion control, reliable delivery, and flow control) in a single
transport layer. In this work, we focus only on connection control,
and argue that this functionality should be logically separate from
data delivery. {\name} can be engineered into an existing transport
protocol like TCP, or into a new sub-transport layer that provides
connection control for multiple data delivery protocols. We give an
example of the latter approach, and share practical lessons on how
{\name} can be integrated in a new network stack~\cite{serval}.


A challenge in realizing an end-to-end protocol like {\name} is that
such protocols are notoriously hard to get right, because of message
re-ordering and subtle corner cases. Mobility and address multiplicity
further exacerbate the problem. To ensure the correctness of {\name}
in face of these challenges, we modeled the protocol in
SPIN~\cite{spin}, formally verifying that it is free from livelocks
and deadlocks---even in face of mobility, packet loss, and
reordering. To our knowledge, this is the first mobility protocol to
be formally verified and the development of the model is one of our
contributions. A unique trait of our model is the inclusion of network
packet loss, duplication, and reordering.  Most previous works on
network verification either did not model message
loss~\cite{ver_no_loss_1} or did not model packet
reordering~\cite{ver_no_reorder_1, ver_no_reorder_2,
  ver_no_reorder_3}.  Fersman and Jonsson~\cite{ver_1} did model
lossy, reordered channels but did not give any details or analysis of
their method of doing so. They also limited their analysis to safety
properties that did not test for livelocks, thus avoiding issues of
fair retransmission of packets (as discussed in
Section~\ref{sec:verification}).

In the process of building the model, we found bugs with both our original
design and with an earlier mobility protocol~\cite{tcp_migrate}.  We used our
verified model to construct a detailed state-transition diagram for the
protocol, which was used as a guide when building our implementation of
{\name}. Our model guarantees that connectivity is preserved in the face of
location dynamism as long as communicating hosts do not move at exactly the same
time.  If hosts do move simultaneously, connectivity can still be preserved if
the network has a special ``redirection middlebox" that temporarily
facilitates the re-establishment of the connection based on the previous
addresses.

This paper makes the following contributions:
\begin{enumerate}
\item Defines correctness requirements for protocols that handle location dynamism 
and path multiplicity.
\item Shows an in-band signaling protocol that meets these correctness requirements
and does not require changing the topological nature of Internet addressing or adding
in-network middleboxes.
\item Formally verifies the correctness of the proposed protocol.
\item Identifies some key points in the design space of protocols of this type. 
\item Implements this protocol as part of a new network architecture~\cite{serval} to serve as a proof-of-concept.
\end{enumerate}

The remainder of this paper is organized as follows.  In Section
\ref{sec:related}, we discuss the requirements that must be fulfilled by a
connection control protocol and analyze the related works to give the reader a
sense of where {\name} fits into the broader network architecture landscape.
Next, we will present the details of the protocol in Section \ref{sec:protocol}
and offer a discussion of the design decisions that went into it.  Section
\ref{sec:use_cases} discusses how this protocol can be used to benefit emerging
technologies.  In Section \ref{sec:verification}, we discuss how we formally
verified that the protocol fulfills its correctness requirements.  Then, we
address the security of the protocol in Section \ref{sec:security} and present
a solution to the problem of simultaneous movement in Section
\ref{sec:simultaneous_movement}.  Finally, we conclude with some final
thoughts.

\section{Protocol Requirements and Related Work}
\label{sec:related}
In this section we define requirements to be met by an end-to-end
connection control protocol to correctly handle both location dynamism
and path multiplicity. We also discuss past works and how they meet
these requirements.

\subsection{Protocol Requirements}

%
%
In the traditional network stack, the transport layer is responsible
for initiating a connection to another end-point, and then taking an
application stream and dividing it into packets to send over the
connection. The network layer delivers these packets according to best
effort, and the transport layer on the other end-point reassembles the
packets into an application stream again. With this division of labor,
the transport layer conflates two separate functionalities:
%
%

\begin{itemize}
\item Data Delivery---takes an application stream and divides it into packets
and flows. It guarantees the delivery semantics required by the application (\eg ordering, reliability)
and handles congestion control.
\item Connection Control---associates flows with the correct network addresses and  
demuxes incoming packets to flows.
\end{itemize}
In this work, we treat connection control and data delivery as
logically separate, focusing on the requirements of connection
control.


Traditionally, protocols used by connection control operate at the
\emph{beginning} and the \emph{end} of a connection, \ie when establishing and
tearing down a flow.  However, to support mobility and changing addresses,
connection control should fulfill the following requirement: \emph{two
communicating hosts are guarantee continued connectivity whenever the network
addresses of either (but not both) host changes or some (but not all) of its
interfaces go down}.  To support this requirement, we develop a connection
control protocol that allows hosts to signal address changes for ongoing flows.
This protocol must operate correctly even when control packets are lost,
reordered, duplicated, or arbitrarily delayed, and must ensure connectivity in
both directions. Further, it should restore connectivity even when changes
happen in quick succession. To handle such cases, the more recent change has to
be able to override a previous one that is no longer valid. For example, when
moving to a new location before completing a (re)connection handshake at a
previous location. 

The {\name} protocol meets the above requirement by allowing
connection control to (re)negotiate the network addresses using
an in-band signaling protocol. The connection control demultiplexes packets to
flows, allowing data delivery to make use of path multiplicity without
dealing with low-level network identifiers.  Because {\name} is
separate from data delivery, it should not concern itself with any
reliability guarantees.

The requirements for connection control explicitly excludes {\em simultaneous
movement} which we formally define as the case when both hosts move before
either one could receive a single packet\footnote{Note that simultaneous
movement talks about a single packet reaching the peer, not about the
completion of a handshake.} from its peer informing it of the peer's new
address.  Therefore, this requirement applies only when at least one of the
hosts is able to successfully inform its correspondent host of the new address
it has acquired. We exclude the case of simultaneous movement because no
in-band signaling protocol can correctly handle this case; rather, additional
techniques must be used, as discussed in Section
\ref{sec:simultaneous_movement}.

\subsection{Related Work}

The problems of location dynamism and path multiplicity have been studied
extensively.  We will first look at alternative approaches to using in-band
signaling.  Then, we will look at other work that has used in-band signaling to
address these issues. 

\subsubsection{Alternative Approaches}

Probably the most widely used solution to location dynamism is Mobile
IP~\cite{mobile_ip, mobileip_ipv6}. This approach uses triangle routing where each
device has a ``home-agent'' with which it registers its current address as it
moves. When a peer wants to reach a particular device, it sends packets to the
device's home-agent, which then forwards the packet to the appropriate
location. The approach has two main drawbacks: (i) it is not as efficient as
in-band signaling since connections need to be established through the home
agent and (ii) it requires a home agent to be aware of a host's location as it
moves, which is a major privacy concern for devices such as cellphones whose
location history mirrors that of the owner.

Other approaches that implement the so-called location/identity
split~\cite{hip,roam,triad,sfr} have sought to change the Internet
architecture to allow addresses to move with devices. The network
became responsible for routing on addresses that were no longer tied
to physical locations. These proposals require significant changes to
the network (impacting deployment) and may not scale well.
\subsubsection{In-band Signaling Protocols}


\begin{table}
\begin{tabular}[c]{l c c c c}
  \bf Feature               & \bf \name  & \bf MPTCP & \bf TCP-Migr  & \bf TCP-R          \\ 
\toprule
  Path Multiplicity           & yes      & yes   & no          & no             \\ 
  Location Dynamism           & yes      & yes   & yes         & yes            \\ 
  Multiple Data-Delivery   & yes      & no    & no          & no             \\ 
  Verified Correct            & yes      & no    & incorrect   & n/a \\ 
  Xmits During Migration  & yes      & yes   & no          & n/a            \\ 
  Tests Reverse Conn.  & yes      & no    & no          & n/a            \\ 
\bottomrule
\end{tabular}
\caption {Comparison of {\name} with related work}
\label{tab:related_work}
\end{table}

A comparison of selected previous works that have addressed either
path multiplicity or location dynamism through in-band signaling is
presented in Table~\ref{tab:related_work}. TCP-R~\cite{tcp-r} was the
first to propose in-band signaling for handling location dynamism but
did not offer any details about protocol operation such as sequencing
or retransmission.  TCP Migrate~\cite{tcp_migrate} specified a full
protocol for in-band signaling to handle migration. However, we have
found that TCP Migrate cannot guarantee reconnections due to corner
cases where packet delay and loss lead to lost connectivity.  This
misbehavior is a result of relying on implicit connection control
acknowledgments through the data stream, as described in
Section~\ref{sec:explicit_ack}. TCP Migrate also gives greater
security guarantees than the current network stack (and {\name}) by
adding protection against on-path hijacking attacks. This added
security, however, comes at the cost of requiring heavyweight
cryptology.

Multipath TCP (MPTCP)~\cite{mptcp} defines a method of using multiple
network paths for one connection at the transport layer.  We envision
that {\name} will be used in conjunction with a transport-layer
protocol like Multipath TCP.  Location dynamism is handled in MPTCP by
starting new flows on new addresses instead of by changing the
addresses associated with existing flows.  In contrast, our approach
allows handling mobility by either starting new flows or moving
existing ones. This added flexibility may be useful to some data
delivery protocols.

%

\section{The {\name} Protocol}
\label{sec:protocol}

\newcommand{\flowidc}{FlowId$_c$}
\newcommand{\flowids}{FlowId$_s$}
\newcommand{\flowidm}{FlowId$_m$}
\newcommand{\flowidcp}{FlowId$'_c$}
\newcommand{\flowidsp}{FlowId$'_s$}

The design of {\name} consists of three main parts.  First, end-points
perform a handshake to establish a connection with a single flow.
Second, the end-points can add more flows to the existing connection
to use additional interfaces or paths.  Third, the end-points can
change the addresses associated with ongoing flows as attachment
points change or interfaces fail. All of these parts are captured in
{\name}'s state machine, as shown in
Figure~\ref{fig:state_machine}. In the rest of the section we will
detail the protocol that moves connections between these
states. Later, in Section~\ref{sec:verification}, we will describe how
we used formal modeling to verify the correctness of the state
machine.

\subsection{Establishing a New Connection With a Single Flow}


{\name} establishes connections and their constituent flows, and
creates the state necessary to map between flows and the underlying
interfaces used for transmission. Each flow is assigned its own
identifier, called a \emph{flowID}, which is essentially an opaque
demultiplexing key that maps packets to socket state. The usage of
flowIDs avoids overloading other identifiers in the traditional
demultiplexing ``five tuple'', thus solving the problem of binding the
flow to a specific combination of IP addresses and ports, which
inhibits mobility.

\begin{figure}[t]
\begin{center}
\includegraphics[scale=0.6]{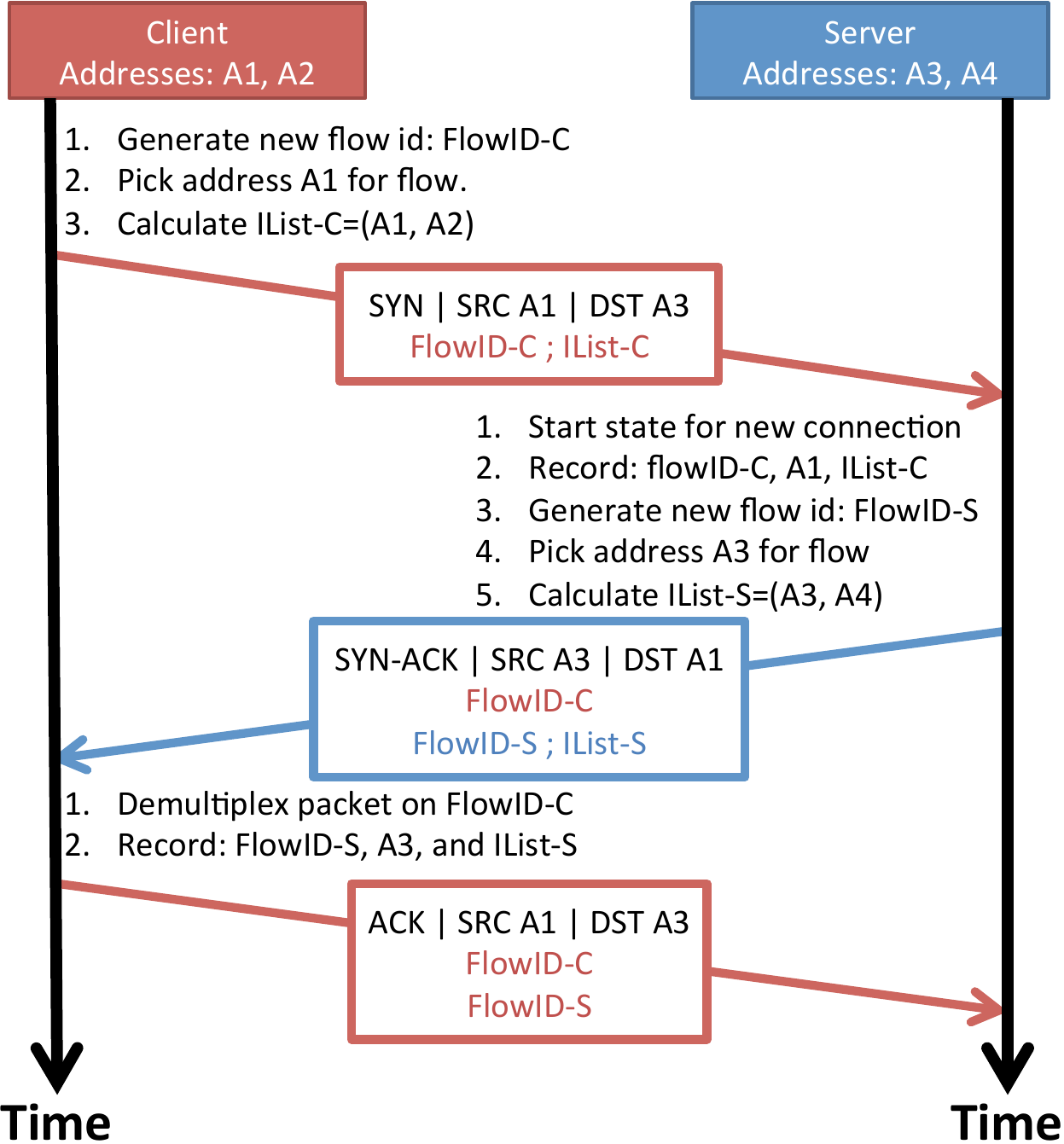} 
\end{center}
\caption{The {\name} protocol for establishing a new connection.} 
\label{fig:protocol-new-connection}
\end{figure}

Connections start with a three-way handshake, as shown in
Figure~\ref{fig:protocol-new-connection}; these messages initialize
the state of the connection and a single initial flow, as shown in
Table~\ref{tab:state}.  After establishing a connection, {\name}
places the appropriate IP address and flowID in outgoing packets and
demultiplexes incoming packets to the right flow based on the
flowID. A list of peer interfaces (ILists) that could be used for
establishing new flows are also exchanged during connection
establishment.  ILists increase connection resilience by allowing for
the establishment of flows on alternative interfaces if the active
interfaces go down.  The connection-establishment protocol ensures
several key properties:

\begin{figure}[t!!]
\begin{center}
\includegraphics[scale=0.6]{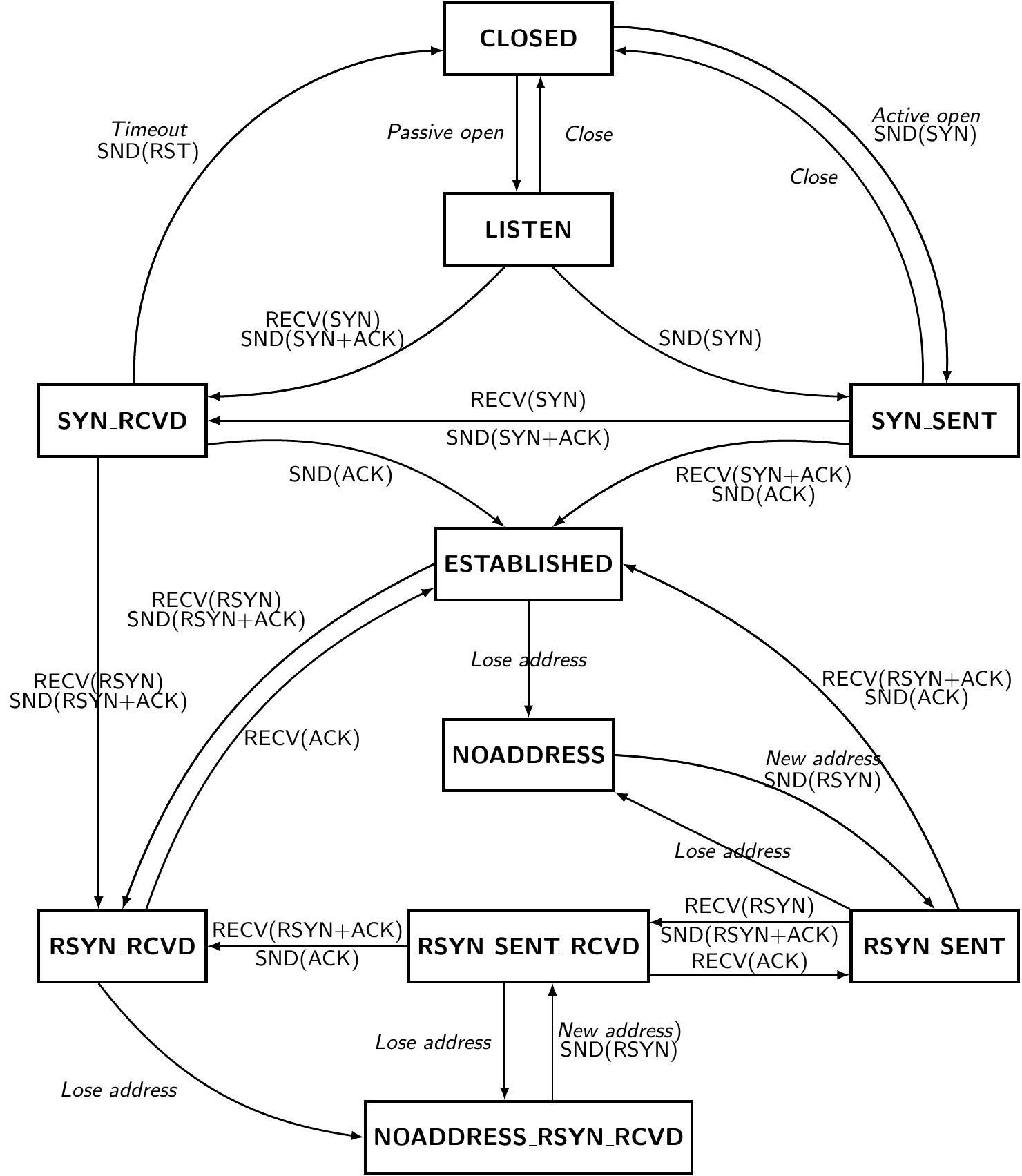}
\end{center}
\caption{The {\name} state machine} 
\label{fig:state_machine}
\end{figure}

\begin{table}
\begin{center}
\begin{tabular}[c]{l l}
\textbf{Abstraction} & \textbf{State} \\ 
\toprule
  Connection & my version number, peer version number \\
             & list of flows, peer interface list (IList) \\ 
\midrule
       & my flowID, peer flowID \\
  Flow & my Address, peer Address \\ 
\bottomrule
\end{tabular}
\end{center}
\vspace{-2ex}
\caption {State stored by {\name} for connections and flows}
\label{tab:state}
\end{table}

{\bf Confirming reverse connectivity.} \label{sec:reverse_connectivity} Network
paths can exhibit asymmetric connectivity, where host A can reach B but B
cannot reach A.  To ensure bidirectional communication, {\name} uses a three-way
handshake where the client sends an ACK to the server to confirm connectivity
on the reverse path, similar to today's TCP.  A similar three-way handshake
is necessary to reestablish connectivity when a flow changes interfaces
or addresses, as discussed in Section~\ref{sec:flow_migration_protocol}.

{\bf Separate demultiplexing keys on each host.}  {\name} uses explicit flowIDs
that uniquely identify the flow. Each flow has two flowIDs, one for each host,
rather than a single shared identifier.  Each host demultiplexes incoming
packets using only its local flowID, but includes the remote flowID in outgoing
packets so the receiving host can demultiplex on its own identifier.  Using two
independent flowIDs offers two main benefits.  First, allowing hosts to pick
their own identifier makes it easier to ensure uniqueness at each end-point.
Second, having separate identifiers simplifies reasoning about the protocol and
proving properties about flow demultiplexing.  

\subsection{Adding Flows to an Existing Connection}

\begin{figure}[t]
\begin{center}
\includegraphics[scale=0.6]{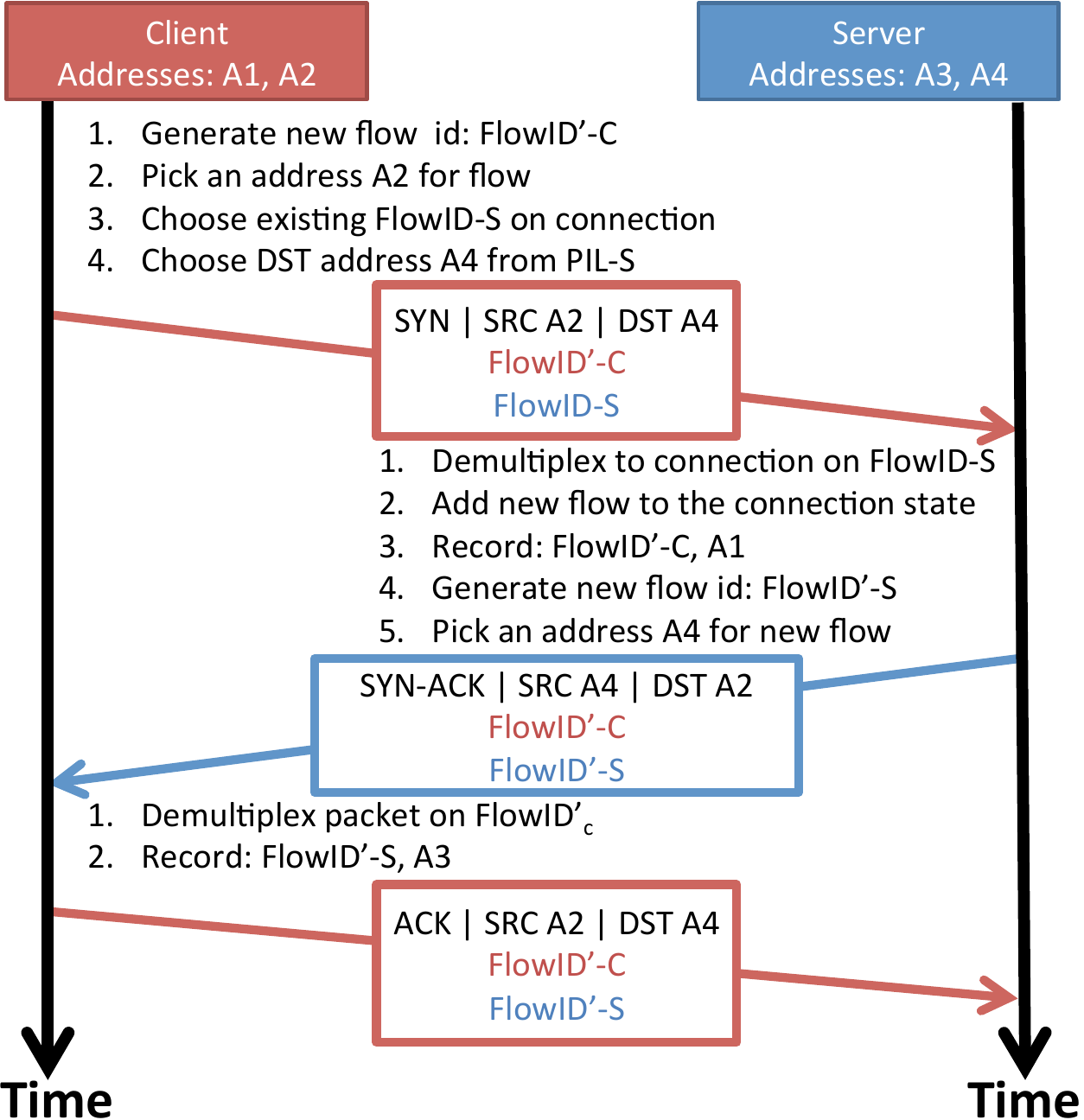} 
\end{center}
\caption{The {\name} protocol for adding a new flow to an existing connection.} 
\label{fig:protocol-add-flow}
\end{figure}

Either end-point can add flows to an existing connection, to spread traffic
over multiple interfaces or paths.  Figure~\ref{fig:protocol-add-flow} shows
how a client adds a flow between local address A2 and server address A4; the
steps for the server to add a flow are analogous.

{\bf Supporting flexible policies for selecting interfaces.}  To establish a
new flow, the two end-points must agree on which pair of interfaces to use.
Each host may have its own policies for selecting interfaces, based on
performance, reliability, and cost.  For example, a smartphone user may prefer
to use a low-cost and high-performing WiFi interface for high-bandwidth
applications, instead of a more reliable (but more expensive) 3G interface. (If
the WiFi connectivity is no longer available, the end-point could migrate the
flow to the 3G interface to continue the connection.)  To support flexible
local policies, {\name} allows each end-point to select its own interface.  The
initiating host selects a local interface (and associated IP address) for the
new flow, and sends a SYN packet to one of the interfaces at the remote
end-point.  Upon receiving the SYN, the remote end-point selects a (possibly
different) interface based on its own local policies, and responds with a
SYN-ACK.  So, while the initiating end-point may influence the decision (\eg by
picking a remote interface based on past performance), the remote end-point has
the final say on which of its local interfaces to use.

\subsection{Changing the IP Addresses of Existing Flows}
\label{sec:flow_migration_protocol}

\begin{figure}[t]
\begin{center}
\includegraphics[scale=0.6]{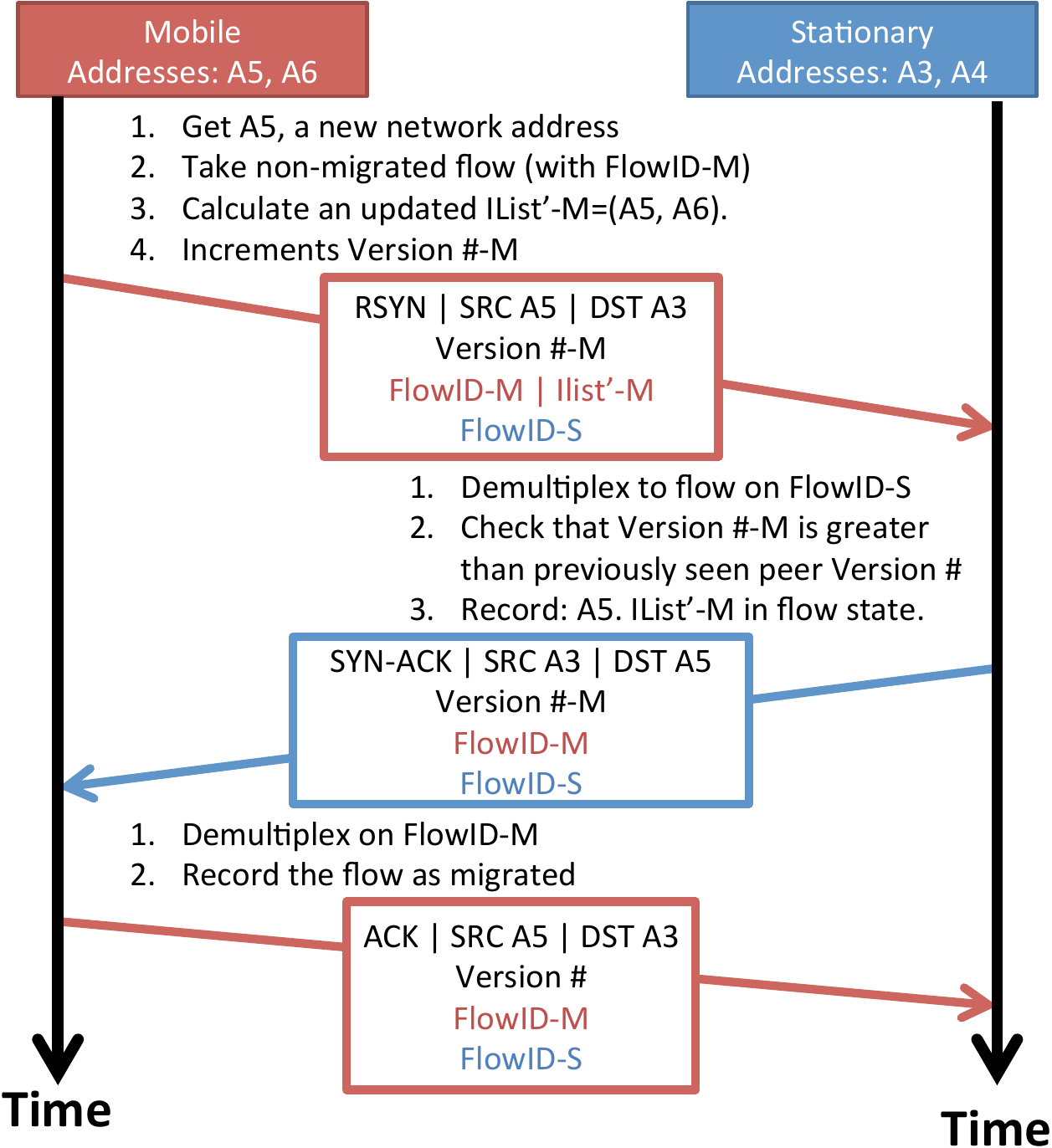} 
\end{center}
\caption{The {\name} protocol for changing the address associated with an
already established flow. {\flowidm} and {\flowids} are the ids for the flow
while A5 is the new address and (A5,A6) is the new interface list.} 
\label{fig:protocol-migrate}
\end{figure}

When a host changes location due to device mobility, VM migration, or failover,
it needs to preserve flow connectivity by notifying its peers of its new
network addresses.  We present the protocol used to update the peers in
Figure~\ref{fig:protocol-migrate}, where the mobile host changes its address
and notifies the stationary host. Once a mobile host establishes a new address
for one of its interfaces, it runs this protocol on every flow using that interface.  

This protocol can also be used to update the IList even if the address on
active flows does not change (\eg an alternative interface established
connectivity).  In that case, the new address on the flow simply remains the
same as the old one; only the IList changes. The IList is always updated as a
single entity with the new list overriding the old one.  No incremental update
protocol is provided to avoid convergence issues. Because the IList is not 
very large, the amount of communication overhead saved with an incremental
update protocol is not worth the added protocol complexity.
 
{\bf Migrating flows independently.}
When a host moves, many of its addresses may change at once. Therefore, it may
seem more efficient to have migration requests refer to connections or to
interfaces instead of having each host migrate each flow independently.
However, this complicates the protocol by introducing dependencies between
flows, which would make the protocol harder to verify. It would also complicate
demultiplexing rules by allowing control messages to affect flows with
different flowIDs.  In addition, disparate flows can have different timeout
timers and different timers related to migration events (\eg how long after a
migration to initiate a migration handshake).  For this reason, all migration
messages concern a single flow.

{\bf Use of version numbers.}
Upon receiving a RSYN packet, a hosts needs to verify that the change of address
requested is not stale.
To this end, the {\name} migration protocol uses version numbers so that a host getting
a new migration request can determine whether the request is newer than the last
one that was processed. For example, if a client moves from Address A1 to A2 to A3, the server
may receive the migration request for A3 before A2, and must then know to
ignore the old migration request to guarantee correctness. We create a new
version number space and do not reuse the sequence space of the transport layer
to make {\name} independent of any given transport level protocol. Furthermore,
a version number is global to the connection and not to an individual
flow to give an ordering to IList updates. This ordering is needed to determine the
most current IList even if it is updated by multiple flows on a connection. 
The initial version number is established
during the initial SYN handshake, much like the initial sequence number in TCP. Version
numbers must increase monotonically across RSYN events. We handle version number
wraparound in the same way as TCP handles sequence number wraparound.

It is noteworthy that our use of version numbers is markedly different than
sequence numbers in TCP.  Namely, we do not use version numbers for reliability.
That is, upon receiving a packet with version number N, we do not verify that
we have previously received all packets with version numbers less than N, but
rather only verify that N is greater than any version number previously
received. This is in stark contrast to sequence numbers, which require the
processing of all packets up to sequence number N-1 before the packet with
sequence number N can be processed.  We use version and not sequence numbers
because at any given time, a host's peer only cares about the current address
on a flow and not the history of address changes that occurred on an interface.

In addition, migration message processing should not be delayed waiting for stale migration
messages, which cannot be processed because the interface addresses have again changed.  
Migration messages should reflect the current state of its interfaces; the history of 
migrations does not matter.  

Originally, we considered an alternative design that avoided version numbers. Since each
change of a version number is accompanied by a three-way handshake, we considered
instead requiring flowIDs to change after each successful migration. This change in flowID
establishes an ordering on migration control packets, as only packets containing
the most current flowID would be accepted as valid.  While this method
appears to work (and we formally verified portions of this protocol), it
adds complexity to the protocol and introduces non-obvious edge-cases
into the demultiplexing rules.  And in the end, it only saved a few bytes of space in control packets. 
Thus, we instead introduced version numbers for a cleaner protocol design.

{\bf Explicit acknowledgments.}
\label{sec:explicit_ack}
\begin{figure}[t]
\begin{center}
\includegraphics[scale=0.6]{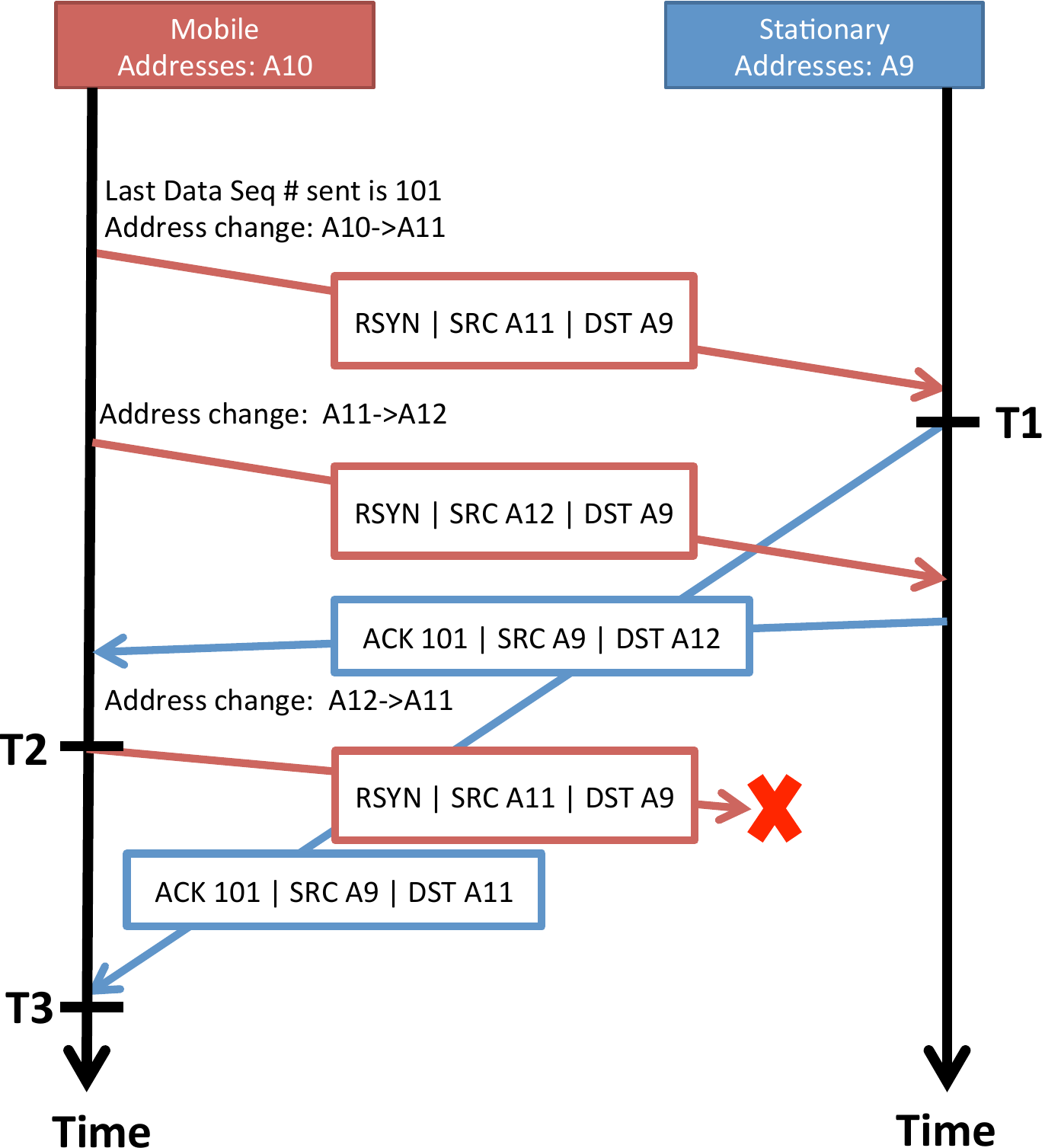} 
\end{center}
\caption{An example of a misbehaving protocol trace when using implicit ACKs. The packet
sent at time T1 is delayed and received at time T3, where it is assumed (incorrectly)
to acknowledge the packet sent at time T2 (which is lost).} 
\label{fig:explicit-ack-counter}
\end{figure}
The {\name} protocol does not use the same method of implicit acknowledgment of
migration messages as TCP Migrate because we found that method to have
misbehaving corner-cases. TCP Migrate used the fact that it received data
packets on the new address as a de-facto acknowledgment that a migration message
was received.  We illustrate one misbehaving corner case in
Figure~\ref{fig:explicit-ack-counter} where the migration at time T2 is lost
but thought to be acknowledged.  To avoid such corner case, we introduce
explicit ACK packets that carry the version number of the migration but can
still be piggy-backed on data packets.

\section{Case Studies: Current and Future Applications}
\label{sec:use_cases}

In this section we will discuss how {\name} can be used to fulfill the full potential of new and emerging technologies. For each technology, we will give a brief overview of its functionality and how
it is limited by the current network stack and then explain how {\name} can help. For some
application we use our implementation of the {\name} protocol to serve as a proof-of-concept
of its utility. The implementation is part of a broader network
architecture project~\cite{serval}, which includes additional changes to the network. But, 
the applications we present here only use the {\name} protocol for their operation. 

\subsection{Mobile Devices} 
Handling physical device movement is a challenge for the current network stack.
Currently device mobility is handled in one of three ways: 
\begin{enumerate}

\item Devices can change IP addresses, break any existing ongoing connections,
and rely on application-level recovery.  

\item Internet providers can use
MobileIP~\cite{mobile_ip} with all of the limitations, as described in
Section~\ref{sec:related}. 

 \item  Providers can use various hacks such as
large layer-2 VLANs to preserve device addresses when the locations of
devices change. This approach gives up scalable address aggregation, making it
feasible only within a single enterprise or campus.  

\end{enumerate}
In contrast, {\name} supports location dynamism by allowing devices to simply
change their addresses as they move.

\subsection{Devices with Multiple Network Interfaces}

\begin{figure}[t]
\begin{center}
\includegraphics[scale=0.8]{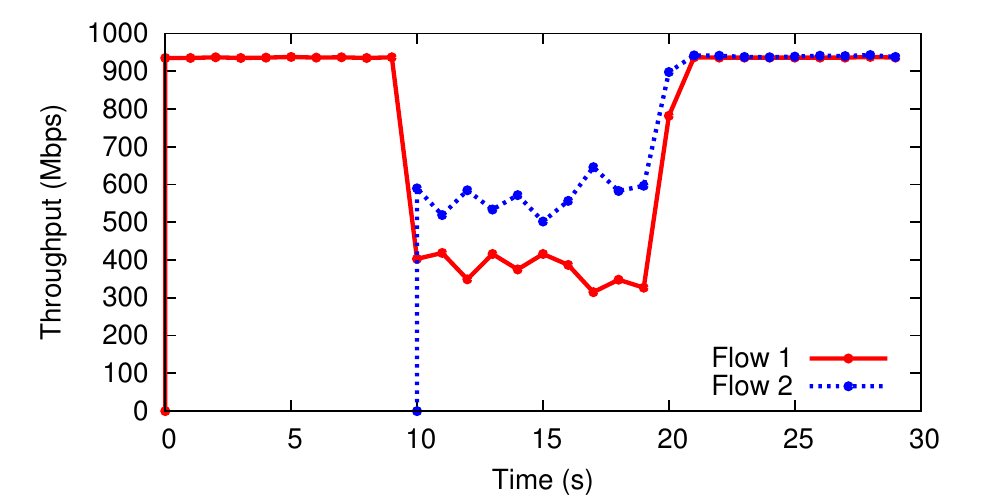} 
\end{center}
\caption{This experiment shows the use of multiple flows in {\name}. Between
time 0s and 10s, only flow 1 is active. At time 10s, flow 2 begins and operates
over the same network paths as flow 1, thus sharing the aggregate throughput. At
time 20s, flow 1's addresses are changed so that it operates over a different 
network path than flow 2, and both flows can then achieve maximum throughput.} 
\label{fig:interface-migration}
\end{figure}

Many Internet enabled devices now have multiple network interfaces. Servers
have multiple network cards, each with its own IP addresses and both smartphones 
and tablets often have both WiFi and 3G connections. These devices,
however, cannot fully leverage this multiplicity of network attachments as
each connection is statically tied to a single network address. Path
multiplicity allows such devices to use all their interfaces for the same
connection.  For example, figure ~\ref{fig:interface-migration} shows how our
implementation can use path multiplicity to get better throughput on a
connection.  At the same time location dynamism allows flows to failover to
alternative interfaces if an interface goes down. 

\subsection{Virtual Machine Migration and Failover}

\begin{figure}[t]
\begin{center}
\includegraphics[scale=0.8]{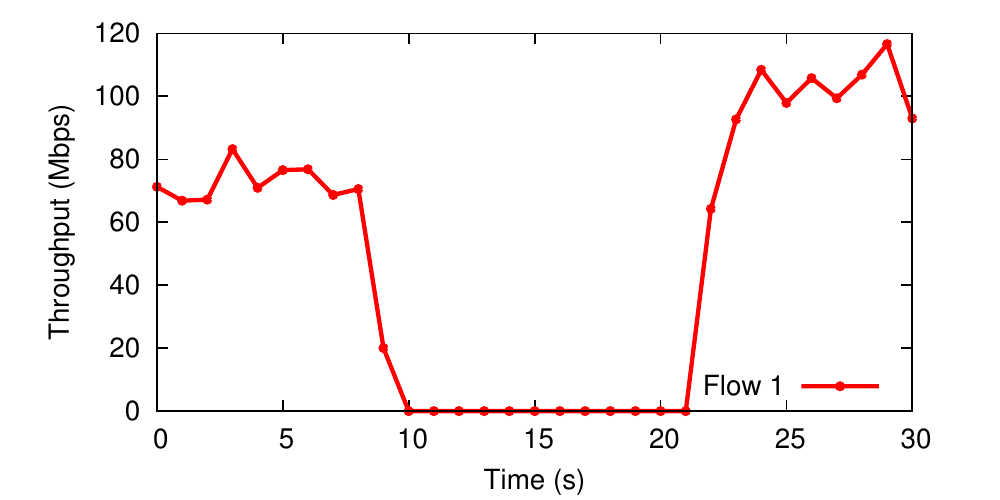} 
\end{center}
\caption{This experiment shows VM migration across layer-2 domains using
{\name}.  The throughput shown is that of a client that communicates with a VM
that migrates at 10s. The gap between 10s and 20s is while the VM has moved to
the new location but not yet received a new address (via the hypervisor cycling
its interfaces, causing a new DHCP request), as well as the delay for TCP to
pickup after retransmission timeouts. This graph shows that communication can
be restored even after a potentially long period of disconnection (although
optimizations for faster address reassignment can be used to reduce this delay
significantly).} 
\label{fig:vm-migration}
\end{figure}

Modern servers often run on virtual machines (VMs) to ease management tasks and
create a more flexible infrastructure.  Recently, two new technologies have
emerged: Live VM Migration and Virtual Machine Failover Replication.  During
live VM migration, a VM on one physical host gets transferred to another
physical host in a way that is seamless for any host communicating with it. In
Virtual Machine Failover Replication~\cite{remus}, a primary VM syncs its state
with a backup VM, which can then act as a hot-spare and take over upon
failure of the primary VM.  

To make these operations appear seamless to hosts communicating with the VM,
ongoing connections need to migrate between physical hosts. Currently, this
requirement limits these operations in that the two machines running the VMs
have to reside on the same layer-2 domain so that the layer-3 address used by
the VM can remain static.  {\name} would allow layer-3 addresses to change
without breaking ongoing connections and thus allow wide-area VM migration and
failover replication.  Figure~\ref{fig:vm-migration} shows the throughput
between a client and a migrating server using our implementation.
 
\subsection{Multi-homing}

Many networks are now serviced by multiple ISPs. This is especially prevalent
in datacenters where a single datacenters can have peering connections with
many providers. Currently, a host has no way to choose the path that its
connections use. A possible solution is to allow ISPs to assign multiple
virtual addresses to a single host interface to allow the host some control
over the possible paths. In this scenario, path multiplicity would allow hosts
to stripe data across different providers for the same connection.

\subsection{Multipath Networking}

The networking community is actively working on ways to support multipath
routing.  However, there has been little research in how to handle
demultiplexing on the end-hosts and allow paths to change in response to
mobility. {\name}'s flow abstraction maps nicely to that of paths by
substituting paths (or pathlets~\cite{pathlet_routing}) wherever we currently refer to addresses.
Path multiplicity would allow multiple paths to be used at once and location
dynamism would allow modifying paths without breaking flows.

\section{Formal Verification} \label{sec:verification}

Distributed protocols such as {\name} are difficult to reason about,
precisely because they involve multiple independent hosts that communicate
asynchronously over unreliable channels.
Hosts execute in arbitrary order, and messages can be lost, reordered, or duplicated.
These factors lead to a large number of possible execution traces of the protocol,
each of which needs to be analyzed for correctness. Analyzing such protocols 
only informally---\eg by considering the most common execution traces---can
lead to a belief in protocol correctness that is later found out to be
false, due to misbehaving edge-cases. Formal analysis, on the other hand, is
hampered by the difficulty in analyzing a very large number of execution traces in 
a timely manner. 

This section discusses our experience analyzing {\name} using
SPIN~\cite{spin}, a formal verification tool that uses a variety of
techniques to cut down on the complexity of the state-space of execution
traces.  Even still, we had to develop several novel approaches for
using SPIN in order to deal with network packet loss and reordering, as
well as to guarantee that packet retransmission timeouts were executed
in a way that did not interfere with protocol liveness verification. Our
model verifies that the {\name} protocol is free from livelocks and
deadlocks, as well as fulfills its correctness requirement.  To our knowledge,
this is the first end-to-end migration protocol to be formally verified.

%
%
In this section, we first discuss the safety and liveness properties
that we use to guarantee {\name}'s correctness, and provide an overview
of SPIN and its verification mechanisms.  Next, we describe our model
for {\name} and the challenges inherent to modeling such networking
protocols.  Then, we discuss the completeness and limitations of our
verification, as well as its results. The full SPIN model is presented
in a technical report~\cite{flexmove:tech}.

\subsection{A Formal Definition of Correctness}

Traditionally, a protocol needs to be verified for two properties to prove
correctness: safety and liveness. The safety property requires that no
execution of the protocol can deadlock.  Deadlocks violate the correctness of
{\name} since connectivity cannot be restored if either host is deadlocked. 

The liveness property verifies that the protocol cannot enter an infinite loop
where each execution of the loop makes no progress towards achieving the goal
of the protocol.  In {\name}, the goal of the protocol is to allow hosts to
communicate with each other, as specified earlier in
Section~\ref{sec:related}.
Thus,
we define the liveness property as the ability to send a message (such as a
ping) to the correspondent host on any flow and get a response back. Verifying
the liveness property guarantees that data can eventually be transferred
between the two hosts on any execution of the protocol.  The combination of the
safety and liveness properties guarantees that our requirement
is satisfied not just for every connection, but for every flow as well. 
    

\subsection{Verification in SPIN}
We now give a very brief overview of SPIN before describing how 
we use it to model {\name}.  SPIN is a C-like language which allows you to
define multiple processes and the communication between them (notably, using
reliable FIFO channels). An execution trace is a single possible execution of the SPIN program
with a particular process execution and message delivery order.  SPIN analyzes all possible 
execution traces to explore all possible protocol executions.

Protocol verification often faces a ``state-space explosion'' problem.
The execution state includes the values of all global variables, local
process variables, and communication queues, defined at a single point in time during an
execution trace. The state-space of the verification refers to the set of all
execution states found in all possible execution traces.  In order for
verification to complete, the state-space must be kept relatively small.
Yet, exploring all possible execution traces of a protocol can easily
create exponential blow-up in the state-space!  One of the biggest challenges
in creating a model is in using the right amount of simplification to avoid
such state-space explosion, while at the same time making sure that the model
remains sound---\ie that these simplifications do not remove misbehaviors that
exist in the real protocol from the model. 

SPIN can perform various checks on the states in the state-space that it
verifies. {\name} uses the following type of checks to verify the protocol:

\begin{itemize}
\item \textbf{Asserts} are those familiar C checks that verify some conditional expression.
These checks are used to sanity-check protocol execution.
\item \textbf{Progress labels} are code labels used to mark pieces of code
that must be executed at least once in any cycle in a execution trace. 
A cycle in an execution trace implies a possible loop in the execution of 
the protocol, and thus needs to be checked for liveness. Progress labels
are a method of specifying liveness properties by requiring some
parts of code to be reached on every iteration of a protocol loop.
There are two progress labels
in this model.  One is located when a host receives a response message back
from its correspondent host. This verifies the liveness property described
above. The second progress label marks the code that models packet loss, as
described in the next section.
\item \textbf{Safety checks} are used to verify that the code never
deadlocks.  They are implemented by simply verifying that each state in the
state-space has a possible transition to another state. This verification
checks that any state visited by SPIN either transitions to another state or
that the state marks the end of one possible run of the verification.  This
verifies that no deadlock exists, since a protocol that is deadlocked would be in
a state that would not be able to transition to any other states.
\end{itemize}
         

\subsection{Modeling {\name} in SPIN}
\label{sec:model-description}

The SPIN verification models two hosts communicating with each other using a
single flow. We first describe how the model represents hosts, communication,
and addresses. Next, we discuss how we represent flowIDs, which present special
challenges due to their randomness. Finally, we describe why modeling only two
hosts and communicating over a single flow is sufficient to prove the
correctness of a protocol that operates in an environment with many hosts and
supports the use multiple flows. 

At a high level, the model represents each host as a different
process. Network communication is modeled using a global array of FIFO queues.
The index of the queue element corresponds to an address. Each host process 
reads from the element in the array corresponding to the address of its
interface and writes to the queue element corresponding to the address
it wants to send a packet to. Modeling migration is done by changing the array 
element a host process uses to receive data. The mobile host then sends {\name}
protocol messages to the stationary host informing it of the new ``address''
(\ie array element) it acquired. The stationary hosts then changes the array
element it uses to communicate with the mobile host. 
We also needed to model {\name}'s randomized flowIDs. But, SPIN, like most
formal method verification methods, cannot deal with randomness well.  In order
to verify a protocol with randomness, the verifier has to evaluate all possible
values for the random variables, which leads to intractable state-space
explosion.  Thankfully, even though the {\name} protocol uses randomness, we
can avoid introducing randomness into the model, while still checking for the
same \emph{semantic} properties in {\name}.  After all, flowID randomness is
used for two purposes: (i) to prevent flowID guessing by off-path entities and
(ii) to ensure that different hosts use different flowIDs. The former property
is a security rather than safety property, and we do not verify security in our
formal model.  The latter property, on the other hand, \emph{is} needed to be
modeled in order to ensure that packets meant for other hosts are dropped (as
discussed previously in Section~\ref{sec:model-description}). But we can avoid
randomness (which prevents flowID collision with high probability) while
preserving unique flowsIDs by just centrally assigning different flowsIDs to
different hosts.  This change allow us to remove the use of randomness in our
model, and therefore avoid the corresponding state space explosion. Note, that
we do not model the highly unlikely case that a flowID collision occurs.  This
case can only affect protocol correctness if it causes a packet meant for one
host to be processed by another. For that to occur, multiple unlikely events
need to happen: a host moves to a new address that was recently occupied by
another host, gets a delayed packet meant for the old host, and that packet has
the same flowID as one of its flows (this event by itself has a probability on
the order of $2^{-32}$).

It is sound to model only two hosts because the protocol insures that hosts
cannot interfere with each other. The only possible way that two hosts could
interfere would be if a packet meant for one host, gets processed by another.
But, two different host would, with high probability, have different flowIDs
for their flows. Since any packet which is received with a flowID that does not
correspond to an active flow is dropped, packets meant for other hosts would
never be processed.

Similarly, it is sound to model only two flows because we can show packets that
are meant for one flow are not processed by another and that shared flow state
can be reasoned about without a formal model.  Incoming packets always get
processed by the correct flow because all demultiplexing is based on flowIDs
which are guaranteed to be unique for each flow on a given host.  The only
shared state that flows (of the same connection) have are peer interface lists
and version numbers, both of which are easy to reason about without formal
models. The only property that is necessary to verify about the correctness of
interface lists is that a host can always update its state with the single most
recent list it got from its peer. This property directly follows from our use
of version numbers.  Since we don't need to model versioning of interface lists
and the only property of version numbers used by the protocol is that they are
monotonically increasing,  we can also ignore the fact that version numbers
are shared across the flows of a connection. From the point of view of an
individual flow, other flows could only cause version numbers to ``skip ahead'',
which would not effect their monotonicity.  

\subsection{Challenges in Modeling an Unreliable Network}

{\name} should operate correctly over a network with only best-effort delivery
guarantees.  Therefore,  our verification has to simulate packet loss,
reordering, and duplication. Modeling these network effects can create
state-space explosion. We now describe how we model these effects in SPIN while
keeping the increase in state-space manageable. Next, we describe challenges
encountered when modeling {\name}'s packet retransmission as a response to
packet loss.

\subsubsection{Loss and Reordering of Network Packets} 

The model of {\name} has to simulate loss and the reordering of packets in the
network.  Natively, SPIN does not model loss and reordering since most
application-layer protocols sit on top of an existing transport layer that
guarantees reliability. {\name}, however, is below the transport layer and
its messages are {\em not} sent reliably.  Previous work~\cite{ver_1} had
identified two major ways of modeling these network effects: (i) a separate
process non-deterministically take packets out of the communication queues and
drop or reorder them and (ii) non-deterministic loss or reordering when sending
or receiving packets. 

After testing both approaches, we conclude that the second approach is much
more efficient. Having a separate process reorder packets leads to more
state-space explosion because the verifier checks all possible interleaving of
the process that simulates packet loss and reorder with the host processes. However, it does not
matter to the protocol {\em when} the packet it received was reordered (\eg five or ten
steps earlier), just {\em whether} a reordering or loss event occurred.  By
limiting loss and reordering events to send and receive operations, we vastly
reduces the state-space without affecting the soundness of the protocol
verification.  The implementation of the operations that simulate the network
effects has to avoid creating unnecessary branches in the state-space.  For
example, even though reordering necessitates the creation of new states
corresponding to the new order of messages in the queue, the changes to the
global state-space should be minimized to this minimal change.  But, reordering
requires the use of temporary variables to store intermediate values.  These
temporary variables are part of the state space and changing their values
creates unnecessary branches in the state space. This is resolved by resetting
all temporary variables to a constant after their use, which merges the
state-space branch back to a common value.

We model network effects inside the send operation. The implementation
of the network effects inside the send operation is hidden from the host, which simply 
invokes the send operation to send packets. This solution encountered some
challenges dealing with the semantics of SPIN. We refer the reader to the technical
report~\cite{flexmove:tech} for further details.

\subsubsection{Timeouts}

Any network protocol that operates over a lossy network needs to have a notion
of timeouts to retransmit packets that may have been lost. SPIN, however, has
no notion of time, and so does not directly model timeouts based on clock time. SPIN
does, however, have a predefined boolean called ``timeout" that is activated
whenever no process can perform any operation.  In effect, the timeout flag
creates a secondary set of operations in each process that are activated
whenever the primary set of operations is blocked for all processes in the
system. In our model, we used this secondary set of operations to perform
retransmission. Intuitively, whenever the regular operation of the protocol
cannot make progress, retransmission kicks in to try to remedy the situation.

The above technique works well if the timeouts that retransmit packets are {\em
fair}. Fairness is a property that states that if we have two or more
processes, each individual process will eventually get a chance to perform its
operations in every execution. Fairness guarantees that a infinite loop
involving only one process will never be explored (\ie all processes are
guaranteed to eventually execute).  This is critical for retransmission
timeouts because the message sent from either one of the host processes could
have been lost, and therefore that particular sender has to retransmit the
packet.  If the retransmission code from the other process is executed
infinitely often, thus starving the sender, then the packet will never be
transmitted and no progress will be made; the verifier will report a progress
violation.  SPIN, however, only has the notion of {\em weak fairness} -- which
means that fairness can only be enforced on operations that can always be
executed.  Our implementation of retransmission---that is, using SPIN's timeout
boolean---does not meet this notion of weak fairness, as it can only be
executed when there are no other actions to take in the system.

Retransmission does not meet the requirements for weak fairness and therefore
SPIN could not natively enforce fairness.
Therefore, we had to explicitly force the model to execute retransmission
timeouts fairly.  Recall that each process in SPIN represents a single host,
each of which may need to perform retransmission of packets to its peer. So, we
enforced fairness among the timeout blocks of all host processes.
This was done by creating a global queue of the host processes and then forcing
the execution of timeouts to occur in the same order as the processes queue.

\subsection{Completeness}

In model checking, the gold standard for verification is if one's model reaches
a {\em fixed point}.  This means that all state transitions from the set of states that have already
been explored lead to other states in this same set. In other words, state exploration
is complete. Unfortunately, this model does not reach a fixed point due to
version numbers. New migration events create new states because they have to
increase the version number and thus new states can always be created. Thus,
this model cannot validate all possible migration events over time. It has,
however, been verified with up to five migrations.  We could not get the model
to verify for 6 migration events due to the increased memory requirements for
the added state-space.  It is believed that all subsequent migrations would be
congruent to the first five, but this has not yet been proven.

\subsection{Results}

The verification of the protocol ran on a Sun SunFire X4100 server with two
dual-core 2.2GHz Opteron 275 processors and 16GB of RAM. As expected, the runtime 
of the verification was highly dependent on the number of migration events that 
could occur. For 5 migration events, the progress property
was verified in 14 minutes and 32 seconds and used 5297 MB of memory; the
safety property verified in 3 minutes 18 seconds and used 3129 MB of memory.  

The model verified the {\name} state machine, which we present in Figure
\ref{fig:state_machine}. An unexpected finding was the RSYN\_SENT\_RCVD state
in the state machine. This state is necessary to ensure correctness when both
hosts move before the migration protocol for either host fully completes.
This state was only found thanks to a progress property violation in a previous
version of the model.

\section{Security}
\label{sec:security}

{\name}, like other connection-based network protocols, is potentially
vulnerable to two main classes of malicious attacks: denial of service (DoS)
and hijacking. A protocol is particularly vulnerable to a DoS attack if a request 
from an unverified party causes a host to spend a 
asymmetric amount of resources.
The classic example of a DoS attack is SYN flooding, where cheaply
crafted (and typically spoofed) SYN packets cause a server to allocate
kernel memory buffers.  Nothing in the {\name} protocol requires
excessive memory or computation to process the initial handshake or the
migration protocol.  SYN cookies can also be used to prevent the
allocation of kernel state to a new connection before return
reachability is tested.

Protocol support for migration introduces new potential threats from
attackers, who may try to (i) hijack ongoing connections by inserting
control messages into the communication stream, or (ii) disrupt
connections by sending fake migration messages.
Fortunately, {\name} prevents such attacks from \emph{off-path} entities
by requiring the presence of nonces during migration.  Nonces are 64-bit
random values that are exchanged during flow setup; all subsequent
control messages, including migration requests, must be accompanying by
the appropriate nonce. Without on-path visibility into the control
messages, off-path entities have no way of determining the correct nonce
without resorting to online brute-force search.  Brute-forcing this
nonce by forging control packets is infeasible, as it will require an
average of $2^{63}$ messages to find a match.  



Migration protocols could also provide protection against \emph{on-path}
attackers.  For example, TCP Migrate~\cite{tcp_migrate} resists on-path
hijacking by using public-key cryptography to secure its control
packets.  On-path entities are still free to simply drop packets, of
course.  {\name} avoids such computationally-expensive means and its
non-cryptographic solution does not mitigate on-path hijacking, but in
this regard, it is no less secure than existing protocols like TCP that
do not support migration.  Connections that require protection against
on-path attackers should use (or are already using) higher-level
mechanisms for securing the data stream, such as SSL.  Securing the data
stream is necessary for data integrity or confidentiality, while neither
{\name} nor TCP Migrate protect against on-path attacks against
availability.

\section{Simultaneous Movement} 
\label{sec:simultaneous_movement}

The {\name} protocol supports mobility whenever the two communicating hosts do
not move at the exact same time. However, an in-band signaling protocol,
without any additional mechanisms, cannot handle simultaneous changes in
location. When two hosts undergo simultaneous movement, each host moves
before it receives a message from its peer about that peer's new address.
In this scenario, each host does not know the new address of its peer. Therefore, 
neither host can notify its peer of its new address and neither will receive a 
notification of its peer's new address. 

We now discuss one mechanism that can enable the communicating hosts to recover
their connection.  It is possible to use a triangle-routing solution which uses
globally-known, statically located, network-level elements (home-agents). In
this solution, each host registers its location with its designated home-agent
as it moves around the network as in Mobile IP~\cite{mobile_ip}. During
simultaneous movement, hosts can contact the home-agents of their correspondent
hosts to learn their new locations. This solution is heavyweight in that it
requires that most hosts on the Internet have static home agents with which
they register their locations, which requires a lot of additional
infrastructure and the purchase of home-agent services.  This solution also
undermines the location privacy of hosts by creating a central location which
is aware of the full movement history of the host. This is an especially big
concern for personal computing devices as we discussed in Section~\ref{sec:related}.  

We propose an alternate solution to allow connection recovery during
simultaneous movement.  In this solution, each network should have a local
redirection middlebox, which keeps a short-lived redirection cache of the new
locations of hosts that have recently moved out of its network. When a host
moves, it should send its new address to the redirection middlebox of its old
network to populate the redirection cache. Upon receiving a new cache entry,
the redirection middlebox takes over (via gratuitous ARP-flooding or a similar
mechanism) the old topological address of the moved host for the duration of
the life of the cache entry. If the redirection middlebox gets an RSYN packet
for an address in the cache, it simply forwards it to the new address of the
host. All packets other than RSYN packets can be dropped by the redirection
middlebox. The address of the redirection middlebox can be learned when a host
joins a network (\eg through DHCP). 

The duration of time during which a redirection box must cache an entry can be
short, measured in seconds, as it just needs to enable a single RSYN exchange
between the two hosts and is not useful after a connection breaks because it
exceeded its retransmission count and timeout. For this to be effective, only
one of the communicating hosts needs to be part of a network with a redirection
middlebox. This scheme is lightweight since the cache entries are short and
decentralized.  It also preserves privacy since a host needs to notify only the
redirection box of the last network it visited of its new address rather than
some central entity that knows the full history of its movements.

\section{Conclusions} 
\label{sec:conclusion}

The Internet architecture needs to evolve to offer better support for
new technologies such as mobile devices and VM migration. Given that a
complete overhaul of the Internet is not realistic, the {\name}
protocol offers a way to incrementally evolve the Internet to support
location dynamism and path multiplicity.  We believe that this
extension to the network stack is relatively easy to deploy and adds
much needed functionality. It can also serve as a robust tool for
future innovation that adds better support for multi-interface and
multi-path communication in the transport layer.

A significant part of this paper was formal verification of the correctness 
properties of the {\name} protocol. This verification was not only useful in
checking the correctness of the final protocol but also motivated the design
by making us aware, early on, of the subtle edge-cases that we needed to consider
for this class of protocols.

{
 \bibliographystyle{abbrvnat}
 \bibliography{local}
}

\end{document}